# Methane as an effective hydrogen source for single-layer graphene synthesis on Cu foil by plasma enhanced chemical vapor deposition†

Yong Seung Kim,[a,b] Jae Hong Lee,[a,b] Young Duck Kim,[c] Sahng-Kyoon Jerng,[a,b] Kisu Joo,[d] Eunho Kim,[a,e] Jongwan Jung,[a,e] Euijoon Yoon,[d,f,g] Yun Daniel Park,[c] Sunae Seo,[a,b] Seung-Hyun Chun[a,b,*]

[a] Graphene Research Institute, Sejong University, Seoul 143-747, Korea

[b] Department of Physics, Sejong University, Seoul 143-747, Korea

[c] Department of Physics and Astronomy, Seoul National University, Seoul 151-747, Korea

[d] Department of Nano Science and Technology, Graduate School of Convergence Science and Technology, Seoul National University, Suwon 443-270, Korea

[e] Institute of Nano and Advanced Materials Engineering, Department of Nano Science and Technology, Sejong University, Seoul 143-747 Korea

[f] Department of Materials Science and Engineering, Seoul National University, Seoul 151-744, Korea

[g] Energy Semiconductor Research Center, Advanced Institutes of Convergence Technology, Suwon 443-270, Korea

* Corresponding author: FAX: +82 2 3408 4316. E-mail address: schun@sejong.ac.kr (S.H.Chun)

## Abstract

A single-layer graphene is synthesized on Cu foil in the absence of $H_2$ flow by plasma enhanced chemical vapor deposition (PECVD). In lieu of an explicit $H_2$ flow, hydrogen species are produced during methane decomposition process into their active species ($CH_{x<4}$), assisted by the plasma. Notably, the early stage of growth depends strongly on the plasma power. The resulting grain size (the nucleation density) has a maximum (minimum) at 50 W and saturates when the plasma power is higher than 120 W because hydrogen partial pressures are effectively tuned by a simple control of the plasma power. Raman spectroscopy and transport measurements show that decomposed methane alone can provide sufficient amount of hydrogen species for high-quality graphene synthesis by PECVD.

Graphene is a two-dimensional carbon material that has been attracting a great interest because of its unique electrical properties and high potential for applications.[1-4] Single-layer graphene has been made by a number of methods, including mechanical exfoliation,[5] epitaxial growth on silicon carbide (SiC) at high temperature,[6,7] and chemical vapor deposition on catalytic metal substrates.[8-12] The most common way of producing single layer graphene is mechanical exfoliation from highly oriented pyrolytic graphite (HOPG).[5] Although the quality of exfoliated graphene is better than others, the dimension of graphene is limited to tens of microns. On the other hand, large-area graphene can be grown epitaxially on single crystalline silicon carbide (SiC) by thermal desorption of Si at the surface. Due to the strong interaction at the graphene/SiC interface, however, the properties of graphene are changed, resulting in band-gap opening at Dirac point.[13] Also, non-uniformity at step edges and high growth temperature (> 1300 °C) limit the application of graphene produced by this method.

The chemical vapor deposition (CVD) method is the most promising approach for producing graphene for large-scale electronic device applications because of its comparative low cost and high efficiency. As the quality of CVD-grown graphene strongly depends on the growth conditions, the underlying growth mechanism of CVD graphene has been studied intensively.[14,15] One of the notable findings is the prominent role of hydrogen in the graphene synthesis revealed by Vlassiouk *et al*.[14] The hydrogen appears to be indispensable as a co-catalyst of the metal in graphene synthesis and as an etching reagent that controls the size and shape of domains. While Gao *et al*. reported the synthesis of graphene on Cu foil in the absence of $H_2$ flow by using extremely elevated methane concentrations,[16] the growth is not self-limited to single-layer graphene in this case.

In contrast, it was recently reported that plasma enhanced CVD (PECVD) could produce single-layer graphene without $H_2$ flow and the quality of graphene was high enough for innovative device applications such as barristor.[17-19] However, the details of growth kinetics are still unknown, and the study on the role of plasma is timely. Here, we present how the plasma affects the nucleation and subsequent growth in the absence of $H_2$ flow. It is found that the plasma plays an important role in the graphene growth by generating hydrogen species during the decomposition process of methane into active species. The amount of hydrogen species increases with the plasma power, resulting in a drastic change of grain size. These behaviors are well explained by the dual role of hydrogen in the graphene growth. When the plasma power is high enough, good-quality graphene with the carrier mobility of ~3,200 $cm^2V^{-1}s^{-1}$ can be produced at 830 °C by using methane as both carbon and hydrogen sources.

The graphene films studied in this work were grown on 25 μm-thick Cu foils (Alfa Aesar, 99.8 %, #13382) by PECVD. The chamber is equipped with an inductively coupled plasma reactor (a schematic diagram of the system is shown in the ESI). The RF coil, having a diameter of ~25 cm, is located 35 cm above the top surface of the graphite substrate holder. In this geometry, the heating effect of substrate by plasma was negligible (see ESI). A residual gas analyzer (RGA 100, Stanford Research System) is attached at the side wall of chamber to investigate the density of discharged species. A differential pumping technique is employed to meet the required working pressure of RGA (detailed information can be found in the ESI). The system is pumped with a turbomolecular pump (Osaka Vacuum LTD, TG1003), keeping the base pressure as low as ~$10^{-7}$ Torr. Polycrystalline Cu foil was cut into 7x7 cm$^2$ pieces and mounted in the chamber without any pre-cleaning treatment. Five different stages were employed to synthesize graphene films on Cu foil using methane (CH$_4$) as a carbon source. The Cu substrate was heated to the growth temperature (700 ~ 830 °C) at the heating rate of 3 °C/s. When it reached the target temperature, H$_2$ gas was introduced into the chamber at the flow rate of 40 standard cubic centimeters per minute (sccm). Hydrogen gas was discharged by an RF power of 50 W for 2 minutes to eliminate surface oxides on the copper foil. Then, the chamber was purged with Ar at the flow rate of 100 sccm for 2 minutes to remove residual hydrogen gas. During the graphene growth stage, radio-frequency (RF) plasma was generated for a specified growth time under a continuous flow of argon (or hydrogen, 40 sccm) and methane (1 sccm), while the pressure was kept at 10 mTorr. The plasma power was varied from 10 to 200 W and the growth time from 0.2 to 4 minutes. Subsequently, the sample was cooled down rapidly to room temperature at a cooling rate of 3 °C/s by turning off the heating power, and then it was taken out for characterization.

All the synthesized graphene films were transferred onto SiO$_2$/Si substrates by etching the Cu foil in an aqueous solution of iron chloride (FeCl$_3$).[8] Prior to wet-etching, the surface of graphene/Cu was spin-coated with poly-methyl methacrylate (PMMA: 950 A2), followed by baking at 150 °C for 3 minutes. The purpose of the PMMA coating is to protect graphene films from cracks and tears during the transfer process.[9, 12] When Cu foil was dissolved completely, the PMMA/graphene membrane was washed with deionized water and placed on the SiO$_2$/Si substrate. Finally the PMMA film was dissolved and removed by acetone.

The surface morphologies of graphene films on Cu substrates are analyzed by scanning electron microscope (SEM: Tescan, VEGA-3) and optical microscopy (Leica, DM2500M). High-resolution transmission electron microscopy (HR-TEM) studies were conducted by JEOL JEM-2100F. The crystallinity and the structural information of synthesized graphene are obtained by micro-Raman spectroscopy (Renishaw, inVia system) using a 514.5 nm laser. In order to estimate the uniformity of synthesized graphene, Raman spectra are measured at three different points in each graphene film. The averaged data of these points are presented in the plots of intensity ratios and the full width at half maximum (FWHM). The point-to-point variation of the Raman spectra is displayed as an error-bar. A minimal laser power of 0.75 mW is used carefully during the measurements to avoid any damage or heating of the graphene films. Electrical properties were measured using the two-probe method.

Methane (CH$_4$) provides hydrogen species during its dissociation process into active species. In thermal CVD, only ~0.0002 % of incoming methane dissociates to active species in the gas phase at a temperature of 900 °C.[20] In PECVD, however, more than 80 % of methane dissociates to other species such as H, H$_2$, CH$_3$, CH$_2$, C$_2$H$_4$, etc. due to the advantage of inductively coupled plasma (ICP).[20-23] The density of discharged species measured by RGA is shown in Fig. 1a. By controlling the plasma power, the density of hydrogen species (H$_2$, H) is tuned effectively; it increases with plasma power and saturates when the plasma power is higher than 100 W (Fig. 1b). Considering the role of hydrogen in graphene growth, it is expected that the early stage of graphene growth will be affected by the hydrogen density. Fig. 2a-c shows SEM images of graphene grains synthesized at plasma powers of 10, 50, and 170 W in Ar/CH$_4$ mixtures. The average grain size is ~0.4 μm when the plasma power is 10 W, but it sharply increases to ~3 μm when the plasma power is increased to 50 W. It also sharply decreased to ~0.8 μm at a plasma power of 170 W. The plots of the grain size and the nucleation density as a function of plasma power in Fig. 2e show that the grain size (the nucleation density) has a maximum (minimum) at 50 W and saturates to ~0.8 μm (~1.8 μm$^{-2}$) when the plasma power is higher than 120 W. This result is due to the increased amount of hydrogen species at higher plasma power. It is well known that the hydrogen strongly affects the resulting grain size because of its dual role in graphene synthesis; an activator of CH$_x$ radicals into more active species (CH$_x$ + H ↔ CH$_{x-1}$ + H$_2$) and an etching reagent of graphene (H + graphene ↔ (graphene - C) + CH$_x$).[14] These two processes compete and affect the growth of graphene depending on the partial pressure of hydrogen. The increase of grain size at low plasma power (10–50 W) is due to the co-catalytic role of hydrogen in graphene growth. On the other hand, the reduced grain size at higher plasma power (> 50 W) can be attributed to the etching effect of the graphene by hydrogen. As a control experiment with H$_2$ flow, we synthesized graphene with H$_2$/CH$_4$ mixtures to compare the resulting grain sizes under various plasma powers. A SEM image of the graphene grains synthesized at plasma power of 50 W is shown in Fig. 2d. The resulting grain size and nucleation density is plotted in Fig. 2f as a function of plasma powers from 20 to 200 W. Interestingly, unlike the graphene synthesis in Ar/CH$_4$ mixtures, the grain size and nucleation density of graphene is not affected by plasma power in H$_2$/CH$_4$ mixtures which show ~0.9 μm and ~1.7 μm$^{-2}$, respectively. This observation implies that enormous amount of hydrogen lead to saturated grain size and nucleation density even at low plasma power. Note that these saturation behaviors are identical to those of graphene synthesized with Ar/CH$_4$ mixtures at higher plasma power (> 120 W), indicating that methane provides sufficient amount of hydrogen species in Ar/CH$_4$ mixtures by the assistance of plasma.

Such plasma power dependent behaviors are also seen in the time required for a full coverage. As the growth time increases, graphene grains continue to grow laterally and coalesce into larger ones, resulting in a fully covering monolayer of graphene on the Cu surface. Fig. 3a shows the SEM image of a continuous graphene film grown under Ar/CH$_4$ condition having wrinkles and non-uniform dark flakes similar to those in thermal CVD-grown graphene.[9, 10] Monolayer graphene is identified by high-resolution TEM measurement on the folded edges of the film (Fig. 3b). When the plasma power is 50 W, the Cu surface is fully covered by graphene in 3 minutes in Ar/CH$_4$ condition and in 1 minute in H$_2$/CH$_4$ condition as shown in Fig. 3c. When the plasma power is increased to 170 W, the full coverage time reduces to 2 minutes in Ar/CH$_4$ condition while no difference is observed in H$_2$/CH$_4$ condition at a plasma power of 150 W. To discuss in detail, we define the coverage rate ($v_{cov}$) and the area growth rate per grain ($v_{grain}$) as the increase of graphene coverage and the increase of individual grain area within unit time $t$, respectively. Since the $v_{cov}$ is related to the $v_{grain}$ and the nucleation density ($n_g$) as $v_{cov} = n_g \times v_{grain}$,[24] we extracted $v_{grain}$ from the results of graphene coverage and nucleation density. Fig. 3d shows the $v_{grain}$ as a function of surface coverage under various plasma powers and gas mixtures. Overall, the $v_{grain}$ decreased when the surface coverage of graphene increased because of the decreased area of the exposed Cu surface which act as a catalyst for the synthesis of graphene.[24] In Ar/CH$_4$ mixtures, the $v_{grain}$ decreased with plasma power and became comparable to that of graphene grown with $H_2/CH_4$ mixtures, indicating that the increased amount of hydrogen species reduces $v_{grain}$ because they etch the graphene. Considering the reduced $v_{grain}$ at higher plasma power, the coverage rate ($v_{cov}$) of graphene on Cu is expected to decrease. However, as the plasma power increased from 50 to 170 W, the $v_{cov}$ increased (Fig. 3c) because the $n_g$ increased to ten times (Fig. 2e) while the $v_{grain}$ reduced to one-third (Fig. 3d).

We further investigated the Raman spectra of graphene synthesized under various plasma powers as shown in Fig. 4a. The three most prominent peaks are the D peak at ~1350 cm$^{-1}$, the G peak at ~1580 cm$^{-1}$, and the 2D peak at ~2680 cm$^{-1}$. The D peak is a defect-induced Raman feature observed in the disordered sample or at the edge of the graphene.[25-28] The G peak is known to be associated with the doubly degenerate phonon mode at the Brillouin zone center, indicating sp$^2$ carbon networks in the sample.[25-28] The low intensity ratio of the D peak to the G peak ($I_D/I_G$) in Fig. 4b indicates a low defect density in synthesized graphene, but slightly increased $I_D/I_G$ at a plasma power of 170 W implies that some structural defects are induced when the plasma power is too high. The 2D peak, which originates from a second-order Raman process, is widely used to determine the thickness of graphene.[25-28] For single-layer graphene, it is well-known that the intensity ratio of the 2D peak to the G peak ($I_{2D}/I_G$) is higher than 2, and the FWHM of the 2D band is close to 30 cm$^{-1}$.[9, 25, 27-30] In our samples synthesized in Ar/CH$_4$ mixtures, the ratio of $I_{2D}/I_G$ and the FWHM of the 2D band are 2.1 and 43 cm$^{-1}$, respectively, when the plasma power is 10 W. But they are highly enhanced to ~3.8 and ~31 cm$^{-1}$ when the plasma power is higher than 30 W (Fig. 4b, c). This enhanced Raman spectra of graphene synthesized at higher plasma power supports that decomposed methane alone can provide sufficient amount of hydrogen species for high-quality graphene synthesis by PECVD. Note that these Raman spectra are close to those of graphene synthesized in $H_2/CH_4$ condition ($I_{2D}/I_G$ ~3.8, FWHM ~29 cm$^{-1}$).

The electrical property of synthesized graphene is evaluated by transport measurement. Graphene films were transferred onto $SiO_2$/Si substrate and fabricated into back–gated transistors by e-beam lithography and $O_2$ plasma etching. The optical image of the device is shown in the inset of Fig. 5. The channel width is 3 μm and the channel lengths are varied from 8 to 50 μm. The electrical resistances of graphene are measured at room temperature by using low-frequency (~17 Hz) lock-in technique. The back gate voltage of ±30 V is applied through the highly doped Si substrate using 300 nm-thick $SiO_2$ layer as the gate dielectric. The dependence of sheet conductance on the applied gate voltages is shown in Fig. 5. The shifted charge neutral Dirac point ($V_{Dirac}$) at ~2 V indicates that there are some extrinsic hole-doping in the graphene film, most likely by residues from etching solutions or PMMA. The carrier mobilities ($\mu$) were extracted by assuming a simple Drude model, $\mu = (ne\rho)^{-1}$, where $n$ is the carrier density, $e$ is the electric charge and $\rho$ is the resistivity.[31] The tunable carrier density ($n$) corresponds to induced charges by the gate voltage, $n = C_g(V_{gate} - V_{Dirac})/e$, where $C_g$ is the gate capacitance of 115 aFμm$^{-2}$ obtained from geometrical considerations.[31] For this sample, we calculated an electron mobility of ~3,200 cm$^2$V$^{-1}$s$^{-1}$ at $n = -1\times10^{12}$ cm$^{-2}$. When it was measured from multiple devices, there were no specific dependence on the channel length (L = 8–50 μm) or the number of grain boundaries crossed. The mobility of graphene could be affected not only by the grain boundaries which act as scattering centers[32] but also by other scattering defects that may arise from the transfer and the fabrication processes.[24, 32] Also, the largest grain size does not guarantee the largest mobility.[33] For comparison, we prepared a single-domain graphene device with mechanically exfoliated graphene from highly ordered pyrolytic graphite (HOPG) and measured its transport properties (the gate voltages dependence of sheet conductance is shown in ESI). The mobility of single-domain graphene with a channel length of 1 μm was found to be ~3,640 $cm^2V^{-1}s^{-1}$, which was comparable to that of our CVD graphene having multiple grain boundaries in the channel. These results suggest that, at least in our works, extra defects other than grain boundaries are more dominant scattering centers limiting the mobility.

The effect of substrate temperature on graphene synthesis in $Ar/CH_4$ condition is investigated in a temperature range from 700 to 830 °C while the plasma power and the synthesis time were maintained at 50 W and 3 minutes, respectively. Fig. 6a shows Raman spectra of graphene film grown at different substrate temperatures. As the temperature decreases, the ratio of $I_D/I_G$ increases from 0.2 to 0.7 and $I_{2D}/I_G$ decreases from 4.0 to 2.2 (Fig. 6b). The higher intensity of the D peak and the broader FWHM of the 2D band (Fig. 6c) at lower growth temperatures suggest that there are gradual increments of disorder for graphene synthesized at lower temperatures. Even at the substrate temperature of 700 °C, however, the ratio of $I_{2D}/I_G$ is higher than 2 and a single-Lorentzian 2D peak is observed, indicating that a monolayer graphene was synthesized with some defects.

In summary, we demonstrated that methane alone can provide sufficient amount of hydrogen species for single-layer graphene synthesis on Cu foil without explicit $H_2$ flow by PECVD. The amount of hydrogen species dissociated from methane can be controlled by the plasma power and affects the nucleation and the grain size of graphene. As the plasma power is increased, the grain size increases at low plasma power (10–50 W) and decreases at higher plasma power (> 50 W), showing a saturation behavior when the plasma power is higher than 120 W. Such plasma power dependent behavior is also seen in the time required for a full coverage which decreases from 3 minutes to 2 minutes when the plasma power is increased from 50 to 170 W. In Raman spectra, the ratio of $I_{2D}/I_G$ and the FWHM of the 2D band are highly enhanced and saturated to ~3.8 and ~31 cm$^{-1}$, respectively, when the plasma power is higher than 30 W. Our work clarifying the dual role of methane (as a hydrogen source as well as a carbon source) helps the understanding of graphene growth mechanism by PECVD and guides further improvements.

## Acknowledgements

This research was supported by the Priority Research Centers Program (2012-0005859), the Basic Science Research Program (2012-0007298, 2012-040278), Nanomaterial Technology Development Program (2012M3A7B4049888), and the Center for Topological Matter in POSTECH (2011-0030786) through the National Research Foundation of Korea (NRF) funded by the Ministry of Education, Science and Technology (MEST). YDK and YDP were supported by the NRF grant funded by the MEST (2012-014248, 2012-0006233). It was also supported by a grant (2011-0031637) from the Center for Advanced Soft Electronics under the Global Frontier Research Program of the Ministry of Education, Science and Technology.

**Notes and references**

**Figures**

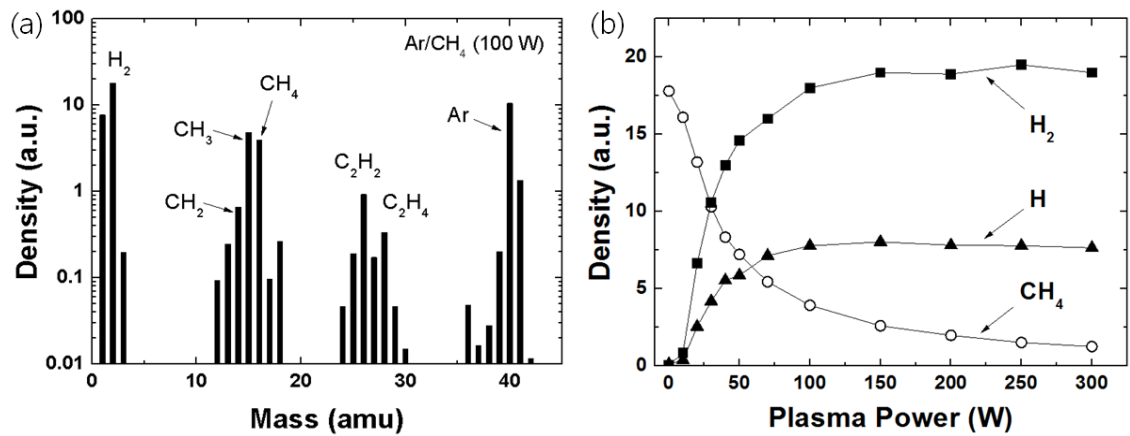

Fig. 1 (a) Mass spectra of the discharge species measured by RGA under a plasma power of 100 W in Ar/CH$_4$ mixtures. (b) Density variations of H$_2$ (filled squares), H (filled triangle) and CH$_4$ (open circles) as a function of plasma power.

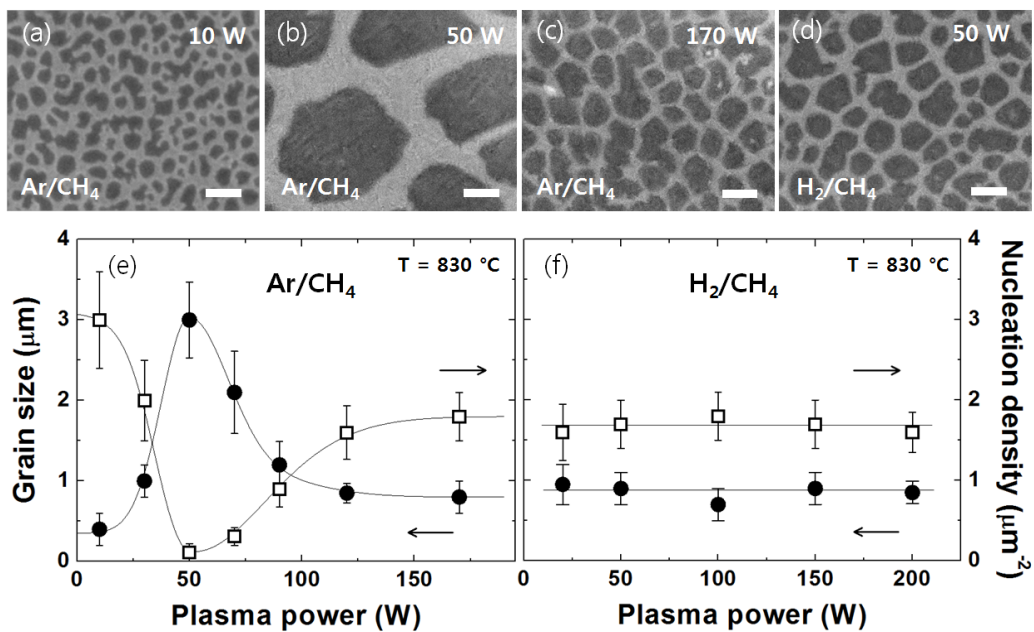

**Fig. 2** SEM images of graphene grains synthesized (a, b) for 1 minute at plasma powers of 10, 50 W and (c) for 20 seconds at 170 W with Ar/CH$_4$ mixtures and (d) for 10 seconds at 50 W with H$_2$/CH$_4$ mixtures. Scale bar = 1 μm. (e, f) Grain size (filled circles) and nucleation density (open squares) of graphene as a function of plasma power under different gas mixtures.

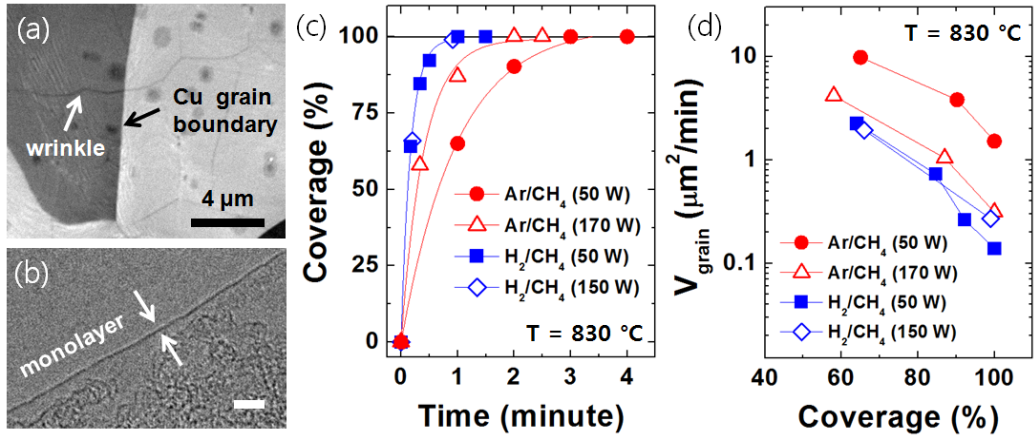

**Fig. 3** (a) SEM image of graphene on Cu foil synthesized at 830 °C for 3 minutes in Ar/CH$_4$ condition with a plasma power of 50 W. (b) High-resolution TEM image of graphene showing the edges of the film containing monolayer graphene film. Scale bar = 2 nm. (c) Graphene coverage on Cu foil as a function of growth time at different gas mixtures and plasma powers. (b) Grain area growth rate as a function of graphene coverage at different gas mixtures and plasma powers.

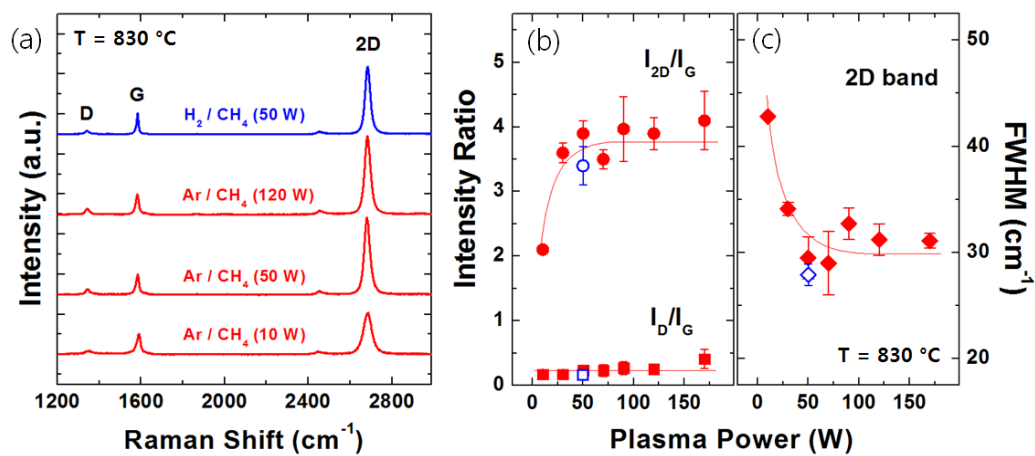

**Fig. 4** (a) Raman spectra of graphene films synthesized with various plasma powers in Ar/CH$_4$ (red) and H$_2$/CH$_4$ (blue) condition. (b) Plasma power dependence of intensity ratio of the D and 2D peak to the G peak in Ar/CH$_4$ (red) and H$_2$/CH$_4$ (blue) condition. (c) FWHM of 2D band obtained from single Lorentzian fit. The lines are guides for the eye.

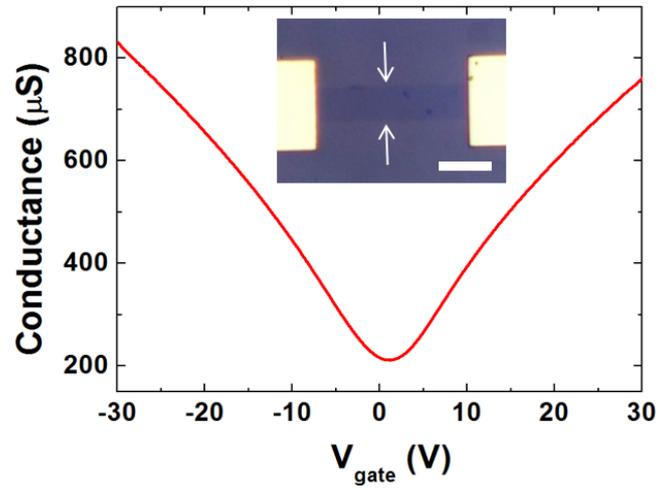

**Fig. 5** Gate voltage dependent conductance of graphene films synthesized for 3 minutes with plasma power of 50 W in Ar/CH$_4$ condition. The inset shows an optical image of FET device. Scale bar = 4 μm.

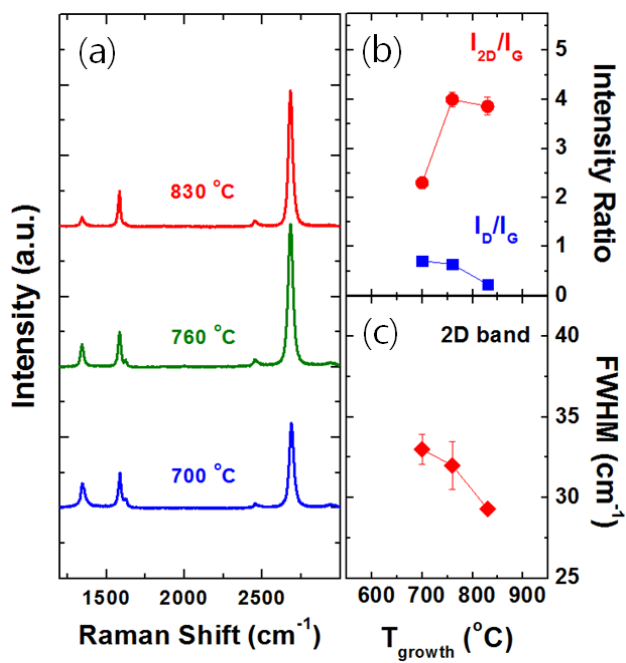

**Fig. 6** (a) Raman spectra of graphene films on SiO$_2$/Si substrate grown for 3 minutes at various temperatures with a plasma power of 50 W in Ar/CH$_4$ mixtures. Temperature dependence of (b) intensity ratios of D and 2D peaks to G peak and (c) FWHM of 2D band of synthesized graphene.

TOC

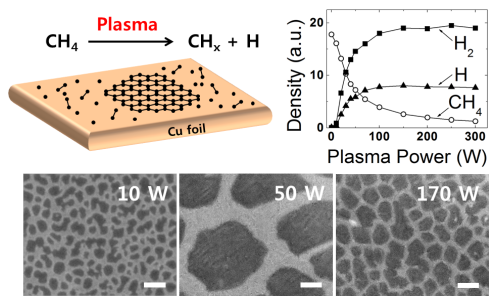

The methane alone can provide sufficient amount of hydrogen species for single-layer graphene synthesis without explicit $H_2$ flow.

# Supplementary Information

Methane as an effective hydrogen source for single-layer graphene synthesis on Cu foil by plasma enhanced chemical vapor deposition

*Yong Seung Kim, Jae Hong Lee, Young Duck Kim, Sahng-Kyoon Jerng, Kisu Joo, Eunho Kim, Jongwan Jung, Euijoon Yoon, Yun Daniel Park, Sunae Seo, Seung-Hyun Chun*[*]

**1. ICP-CVD system**

1.1 Detailed geometry

Fig. S1a shows a schematic diagram of the inductively coupled plasma chemical vapor deposition (ICP-CVD) system. At the top of the chamber, RF coil is mounted below the inlet ports which allow gas flow into the chamber. The diameter of RF coil is about 25 cm and the center of the coil is located 35 cm above the top surface of the graphite substrate holder. Plasma is generated by applying 13.56 MHz RF bias power to the RF coil. No bias voltage is applied to the substrate stage, because it is known that the electric field aligned perpendicular to the surface prevents planar growth of graphene and results in vertically standing graphene sheets.[1,2]

1.2 Substrate heating by plasma

To investigate the heating effect of substrate by RF plasma, the substrate temperature is monitored under different plasma powers of 50 and 400 W. It is conducted without filament heating of substrate and the temperature is measured by a K-type thermocouple which is attached to substrate. When the plasma power is 400 W, the temperature increases only about 3 °C after 12 minutes (Fig. S1b). When the plasma power is reduced to 50 W, the temperature does not change even after 14 minutes. Considering the plasma powers (< 200 W) and the growth times (≤ 3 minutes) used in the graphene growth, we conclude that the substrate heating effect by RF plasma is negligible in our system.

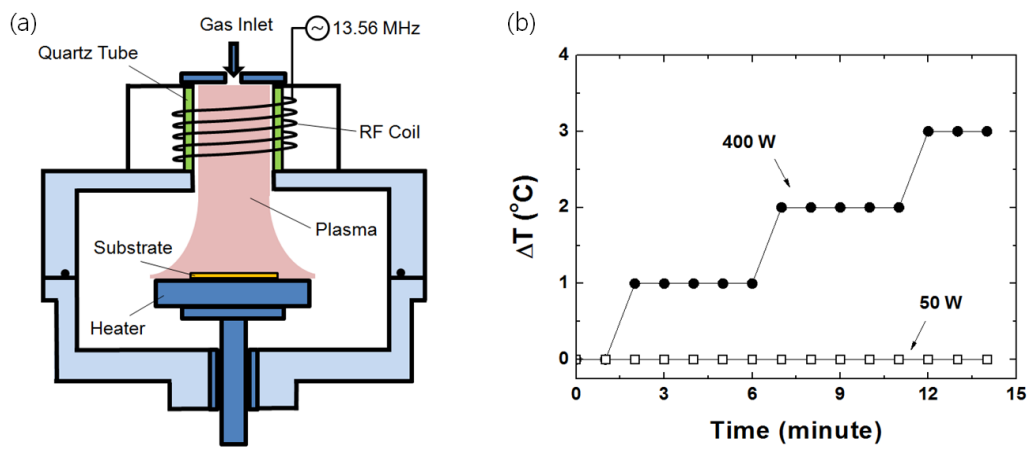

**Fig. S1** (a) Schematic diagram of ICP-CVD system. Remote plasma is generated above the substrate. (b) Substrate heating effect by plasma under different plasma powers of 50 and 400 W.

**2. Differential-pumping technique for mass spectra measurement**

A residual gas analyzer (RGA 100, Stanford Research Systems) is attached at the main chamber to investigate the mass spectra of discharged species. The high pressure of ~$10^{-2}$ Torr during the graphene growth is not suitable for the measurement because the RGA system requires a high vacuum (<$10^{-4}$ Torr) environment. While the pressure could be lowered to the desired level by reducing the gas flow, plasma could not be generated in this case; the pressure should be higher than $10^{-3}$ Torr to generate plasma. To overcome these two conflicting issues, we employed a differential pumping technique[3] in our ICP-CVD system by mounting an aluminum aperture (inner hole diameter ~5 mm) underneath the plasma reactor as shown in Fig. S2. Due to the highly reduced pumping conductance between the plasma reactor and the main chamber, we are able to keep ~$10^{-2}$ Torr in the plasma reactor and ~$10^{-4}$ Torr in the main chamber under proper gas flows (Ar/$CH_4$ = 5:5 sccm). We investigated the mass spectra of discharged species in this configuration.

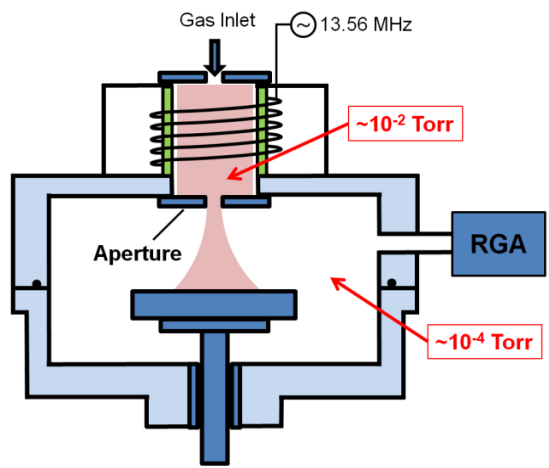

**Fig. S2** Illustration of ICP-CVD system with an aluminum aperture mounted underneath the plasma reactor for differential pumping.

## 3. Transport properties of a single-domain graphene device

A single-domain graphene device is fabricated using mechanically exfoliated graphene from HOPG. The channel length and width of this device are 1 and 3 μm, respectively. The dependence of sheet conductance on the applied gate voltages is shown in Fig. S3. We calculated an electron mobility of ~3,640 cm$^2$V$^{-1}$s$^{-1}$ at $n = -1 \times 10^{12}$ cm$^{-2}$.

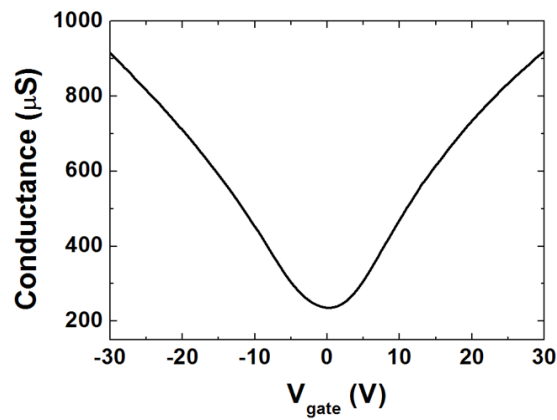

**Fig. S3** Gate voltage dependent conductance of exfoliated graphene films from HOPG.